# Superchiral photons unveil magnetic circular dichroism

S. W. Lovesey[1,2], J. T. Collins[3], S. P. Collins[1]

1. Diamond Light Source, Chilton, Oxfordshire OX11 0DE, UK
2. ISIS Facility, STFC, Chilton, Oxfordshire OX11 0QX, UK
3. Centre for Photonics and Photonic Materials, Department of Physics, University of Bath, Bath BA2 7AY, UK

**Abstract**. Polarization-dependent photon spectroscopy (dichroism) using signal-enhancing superchiral beams is shown to be sensitive to magnetic properties of the sample, whereas previous investigations explored charge-like electronic properties of chiral samples. In the process of unveiling the potential to observe magnetic circular dichroism (MCD), we underline an affinity between spectroscopies using the Borrmann effect, twisted beams and superchiral beams. Use of an effective wavevector in a quantum-mechanical theory unites the aforementioned spectroscopies and vastly improves our understanding of their advantages. Exploiting an effective wavevector for superchiral beams, natural circular dichroism (NCD) is derived from electric dipole - magnetic dipole (E1-M1) and electric dipole - electric quadrupole (E1-E2) absorption events, and MCD is derived from electric quadrupole-electric quadrupole (E2-E2) absorption. Signal enhancement by superchiral beams is a straightforward gain for the user because NCD and MCD are otherwise precisely the same as for a circularly polarized beam, according to our calculations. Our analysis shows that enhancement of E2-E2 is superior to that available for parity-odd events under consideration. Electronic degrees of freedom in all dichroic signals are encapsulated in atomic multipoles that are frequently used in theoretical interpretations of several established experimental techniques.

## I. INTRODUCTION.

Superchiral light can retrieve electronic properties of matter that are invisible with traditional, classical spectroscopies [1-5]. By generating precisely sculpted optical beams, coupling to chiral properties - ubiquitous in biomolecules - can be enhanced dramatically, e.g., Pellegrini *et al*. report a hundred-fold enhancement in natural circular dichroism (NCD) [3]. Our studies use the standing-wave field introduced by Tang and Cohen [2], with two beams related by mirror reflection. Notably, the corresponding spectroscopies using E1-M1 and E1-E2 absorption for NCD and E2-E2 absorption for magnetic circular dichroism (MCD) are found to be intimately related to Borrmann spectroscopy, where we previously demonstrated that suppression of dipole-allowed transitions creates a very strong enhancement of E2-E2 absorption [6, 7, 8]. Enhancement of NCD and MCD by a superchiral standing wave-field is likewise the gift of cancelling dipole-allowed transitions [1, 2].

Photon spectroscopy presents information on electronic degrees of freedom at an atomic level of detail. For the majority of applications, in chemistry, life-sciences and physics, electric dipole-allowed absorption (E1-E1) is the engine of success in unravelling complex and frequently enigmatic, phenomena. Intrinsically weaker signals can be intentionally selected in the preparation of an experiment for their unique electronic information. Specifically, the Borrmann effect and twisted photon beams enhance and isolate electric quadrupole absorption (E2-E2). Building on a previous investigation that unified dichroic signals retrieved with the Borrmann effect and twisted photon beams [9] we are able to adopt superchiral beams in a family of photon-electron spectroscopies. A judiciously chosen effective wavevector in an established quantum-mechanical formulation of dichroic signals provides the familial linkage.

The Borrmann effect, or thick crystal Laue diffraction, arises in high-quality crystals with simple chemical structures, where interference between the incident and diffracted beams sets up a standing wave-field perpendicular to the crystal planes with polarization in the plane of the diffracting atoms. The electric field at the atomic planes vanishes, cancelling dipole-allowed absorption and permitting 'anomalous transmission'. However, while the field intensity vanishes, the field gradient (now perpendicular to the atomic planes) persists and leads to strong quadrupole absorption (E2-E2) that dominates the recorded spectra, in the absence of significant vibration of the diffracting ions. In a number of experimental reports [6, 7, 8], huge quadrupole enhancements at x-ray K- and L-edges are shown, along with a strong dependence on temperature. Enhancement offered by a superchiral beam is very similar to that of the Borrmann effect with absorbing ions at the nodes of a standing wave. Indeed, we note that superchiral beams can likewise be generated by Bragg diffraction within high-quality crystals. The superchiral standing wave-field is formed by two oppositely-directed circularly polarized beams and it possesses spatial chirality when the amplitudes of the beams are unequal [1, 2]. Helicities of the two beams are opposite (as would be realised by a mirror reflection), but the electric vector is the same for both beams due to the reversal of both the helicity and wavevector. The most significant departure from the Borrmann case is that there is no requirement for the photon beam to be linearly polarized and so the case of circular polarization can be considered.

We focus on general principles for (bulk) optical signals in the present communication. Borrmann spectroscopy is revisited in the next section as an introduction to the cancellation of dipole-allowed absorption, and use of an effective wavevector for suitable sculpted photon beams. Thereafter, we use an effective wavevector approach to calculate energy-integrated absorption spectra, derived from the Kramers-Heisenberg dispersion formula, and we call these quantities dichroic signals. The sculpted beam used in our calculations possesses perfect circular polarization [2]. Sec. IV is concerned with NCD derived from one or other of two absorption events, E1-M1 and E1-E2, chosen because they are likely the most useful in applications of the technique to chiral samples. Unlike previous studies of NCD retrieved from superchiral beams, our dichroic signals are expressed in terms of atomic multipoles. This scheme of working provides multiple advantages that include easy access to measurements with alternative techniques, and the facility to readily incorporate symmetries in the sample, e.g., symmetry imposed by a space group. Magnetic materials in the context of superchiral beams have not been studied previously. Our calculation of MCD utilizing an E2-E2 absorption event is reported in Sec. V. Atomic multipoles that arise in MCD are discussed in Appendix A for the sake of completeness. Enhancement factors for dichroic signals offered by superchiral beams are presented in Appendix B. Sec. VI is a brief discussion of our principle findings.

## II. PREAMBLE

By way of an orientation to our calculations in the next sections we present the case for the suppression of dipole-allowed transitions in the Borrmann effect. The standard electric multipole expansion of the electron-photon interaction, V, treats the product of the electron position (**r**) and photon wavevector (**q**) as a small quantity, i.e., (**r**•**q**) << 1 [10, 11]. Absorption that attends the Borrmann effect is similar with two modifications: first, the electric field for a single travelling wave is replaced by that of two waves with one along the primary beam direction, **q**, and the other, of equal amplitude, along the secondary beam direction, **q**′, where the two waves are phased so as to give zero-field at the diffracting planes; second, absorption is by ions displaced from their ideal positions by a small distance **u**. The resulting form of V for the Borrmann effect is then,

$$V \propto \mathbf{r} \cdot \boldsymbol{\varepsilon} \left[\exp\{i(\mathbf{r}+\mathbf{u}) \cdot \mathbf{q}\} - \exp[\{i(\mathbf{r}+\mathbf{u}) \cdot \mathbf{q}'\}]\right] = i\mathbf{r} \cdot \boldsymbol{\varepsilon} \left[(\mathbf{q}-\mathbf{q}') \cdot (\mathbf{r}+\mathbf{u})\right] + \ldots \quad (1)$$

The polarization vector **ε** obeys **q**•**ε** = **q**′•**ε** = 0. The difference between the two terms in V is by design of the relative phase. In consequence, the leading (dipole) term in the normal absorption case has vanished, leaving two terms of the same order. The first of these is a quadrupole term [**r**•**ε** (**q** − **q**′)•**r**] that presents E2 absorption through a term quadratic in electronic positions, and the second is dipolar absorption linear in **r** that is not of immediate interest [6, 7, 8].

To complete a mapping of the Borrmann effect to the standard E2-E2 scattering amplitude ∝ V$^+$V we use Cartesian coordinates in Fig. 1 (top panel), with our visualization of wavevectors for the effect depicted in Fig. 1 (bottom panel). Appendix A contains additional information on the scattering amplitude. Cartesian forms of the wavevectors in Fig. 1 are **q** = q(0, −sinθ, cosθ) and **q**′ = q(0, sinθ, cosθ) which leaves (**q** − **q**′) parallel to the y-axis. According to (1) it is possible to formulate the Borrmann case with the normal calculation of E2-E2 absorption by substitution of an effective wavevector proportional to (**q** − **q**′), given that **ε** is common to both beams. One can proceed with **κ** = (0, 1, 0) as an effective wavevector. In addition, the product of **ε** and **κ** in [**r**•**ε κ**•**r**] is separated from electronic variables to form atomic multipoles that encapsulate degrees of freedom in the ground state. Quadratic electronic positions encountered in an E2 event can be expressed as a sum of a scalar (rank 0), a dipole (rank 1) and a quadrupole (rank 2), by application of the triangle rule for the product of dipoles. Likewise, electronic positions in the E2-E2 absorption event may be represented by a sum of tenors with rank $K$ = 0, ... , 4. The formulation we describe for the Borrmann effect has been applied also to dichroic signals obtained with twisted photon beams [9], and it is exploited in this communication for signals obtained with a superchiral beam.

## III. SUPERCHIRAL BEAM

We return to Fig. 1 and consider two near oppositely-directed beams realized by $2\theta \approx 180°$. A polarization vector is common to both beams so we can exploit an effective wavevector once again. As in the Borrmann case discussed in the previous section, the effective wavevector is parallel to the y-axis of Fig. 1. However, polarization vectors lie in the x-z plane, to a good approximation (paraxial approximation). Results from subsequent calculations for superchiral beams are likely easier to assimilate when our Cartesian coordinates are changed so that the primary and back-scattered beams propagate parallel to a new z-axis and polarization vectors lie in a new x-y plane, and these axes are depicted in Fig. 2 for clarity. The effective vector $\boldsymbol{\kappa} = (0, 0, 1)$ with spherical components $\kappa_{\pm 1} = 0$, $\kappa_0 = 1$ ($R_{+1} = -(x + iy)/\sqrt{2}$, $R_{-1} = (x - iy)/\sqrt{2}$, $R_0 = z$ for a vector $(x, y, z) = \mathbf{R}$). Circular polarization represented by a polarization vector $\boldsymbol{\varepsilon} = (1, i, 0)/\sqrt{2}$ is right-handed and corresponds to a Stokes parameter $P_2 = +1$ (definitions of Stokes parameters follows references [9, 11]).

A superchiral standing wave formed from circularly polarized counter-propagating beams is depicted in Fig. 3, with additional representations in Supplementary Material. It is worth stressing that superchiral signal enhancement arises when the counter-propagating beam has slightly different amplitude to the forward beam [1, 2]. Appendix B contains a discussion of the dissymmetry factor for NCD using E1-M1 and E1-E2 absorption events. It vividly demonstrates a close connection between our approach to properties of superchiral beams, outlined in the foregoing paragraph, and that of Tang and Cohen [1, 2]. In particular, that excess in dissymmetry, or enhancement, is caused by suppression of the E1-E1 event as in the Borrmann effect we introduced in Sec. II. Our reasoning in Appendix B extends to the E2-E2 event.

## IV. NATURAL CIRCULAR DICHROISM

Previous work on dichroic signals derived from superchiral beams has focused principally on NCD obtained with the E1-E2 absorption event, and we start with this case. Our formulation for conventional applications of the E1-E2 event is described in references [11, 12], and we adopt the same notation and definitions as far as it is reasonable to do so. The product of $\boldsymbol{\varepsilon}$ and $\boldsymbol{\kappa}$ in the E2 absorption amplitude is written,

$$h(r) = \{\varepsilon \otimes \kappa\}^2_r = \sum_{\alpha,\beta} \varepsilon_\alpha \kappa_\beta (1\alpha\ 1\beta\,|\,2r). \qquad (2)$$

Here, the symbol $\{\varepsilon \otimes \kappa\}^2_r$ denotes a tensor product of $\boldsymbol{\varepsilon}$ and $\boldsymbol{\kappa}$. The quantity $(1\alpha\ 1\beta\,|\,2r)$ in the second equality is a standard Clebsch-Gordan coefficient created for two dipoles to produce a spherical tensor of rank 2, and it can be different from zero when projections obey $\alpha + \beta = r$. In the same spirit, the E1-E2 amplitude is expressed in terms of,

$$\tilde{N}^K_Q = (i/\sqrt{5})\,\{h \otimes \varepsilon'\}^K_Q \text{ and } N^K_Q = (i/\sqrt{5})\,\{h' \otimes \varepsilon\}^K_Q, \qquad (3)$$

where $\boldsymbol{\varepsilon}' = (1, -i, 0)/\sqrt{2}$, and $h'(s)$ required in $N^K_Q$ are values of (2) evaluated with this polarization vector (factors $(i/\sqrt{5})$ in (3) keep us aligned with expressions in references [11, 12]). Projections obey $-K \leq Q \leq K$, and values $K = 1, 2$ & $3$ are enforced by the Clebsch-Gordan coefficient $(2r\ 1\alpha | KQ)$. The amplitude F(1, 2) for NCD using the E1-E2 absorption event is [11, 12],

$$F(1, 2) = \sum_{K,Q} i^{K-1} (-1)^Q \{\tilde{N} - N\}^K_{-Q} \Psi^K_Q(U), \qquad (4)$$

with,

$$\Psi^K_Q(U) = \sum_{\mathbf{d}} \langle U^K_Q \rangle_{\mathbf{d}}, \qquad (5)$$

and the sum is over all ions identified by the selected atomic resonance. We say more about time-even polar multipoles $\langle U^K_Q \rangle$ at the end of the section. Time-odd (magnetic) multipoles make no contribution to F(1, 2), and, likewise, no contribution to the amplitude using an E1-M1 absorption event.

Evaluating $\tilde{N}^K_Q$ and $N^K_Q$ with values of $\boldsymbol{\varepsilon}$, $\boldsymbol{\varepsilon}'$ and $\boldsymbol{\kappa}$ appropriate for a superchiral beam yields a NCD amplitude,

$$F(1, 2) = -(1/\sqrt{5})\, P_2\, \Psi^2_0(U). \qquad (6)$$

Our conclusion is that NCD from an E1-E2 event detected by a superchiral beam is related to polar quadrupoles $\langle U^2_0 \rangle$ formed by valence electrons of resonant ions.

NCD from an E1-M1 event reveals monopoles and quadrupoles. A derivation of the corresponding amplitude F(1, 1) is similar in many ways to the foregoing work for an E1-E2 event. Following the derivation of the E1-M1 scattering amplitude in reference [13],

$$F(1, 1) = \sqrt{(2/3)}\, P_2\, [\sqrt{2}\, \Psi^0_0(U) - \Psi^2_0(U)], \qquad (7)$$

for a superchiral beam. A polar monopole $\langle U^0_0 \rangle$ and the Stokes parameter $P_2$ are pseudo-scalars.

Angular brackets in the atomic multipole $\langle U^K_Q \rangle$ denote an expectation value, or time-average, of the enclosed tensor operator of rank $K$, i.e., atomic multipoles depend on the electronic ground-state. The discrete symmetries of $\langle \mathbf{U}^K \rangle$ are parity and time-reversal, with multipoles parity-odd (polar) for E1-E2 and E1-M1 absorption events and time-even, or charge-like. Multipoles are Hermitian with a complex conjugate $\langle U^K_Q \rangle^* = (-1)^Q \langle U^K_{-Q} \rangle$, so $\langle U^K_0 \rangle$ is purely real.

The monopole $\langle U^0_0 \rangle$ can be represented by $\langle \mathbf{S} \cdot (\mathbf{S} \times \mathbf{r}) \rangle = i \langle \mathbf{S} \cdot \mathbf{r} \rangle$, where $\mathbf{S}$ and $\mathbf{r}$ are electronic spin and position operators. (Photon helicity $\equiv P_2$ is the average value of $\boldsymbol{\sigma} \cdot \mathbf{p}$, where σ and **p** are spin and linear momentum, respectively. Reflection in a mirror reverses helicity since σ is unchanged by a mirror while **p** is reversed by a mirror.) A tensor product $\{\mathbf{S} \otimes \mathbf{\Omega}_L\}^2_0$ = $[3S_0 \Omega_{L,0} - (\mathbf{S} \cdot \mathbf{\Omega}_L)]/\sqrt{6}$ formed with an orbital anapole $\mathbf{\Omega}_L = [\mathbf{L} \times \mathbf{r} - \mathbf{r} \times \mathbf{L}]$ can be used as an equivalent operator for a polar quadrupole $\langle U^2_0 \rangle$. All multipoles for resonant absorption are functions of the total angular momentum for the core state $J_c = (l_c \pm 1/2)$. So-called sum-rules relate integrated intensities at the two edges, with $J_c = (l_c - 1/2)$ and $J_c = (l_c + 1/2)$, to specific quantities in the ground-state; sum-rules for E1-E2 and E1-M1 absorption events are in references [11, 12, 13]. An Appendix to this communication considers sum-rules for MCD derived from an E2-E2 event, which is the subject of Sec. V.

The electronic factor $\Psi^K_Q(U)$ can be different from zero for electronic structure that is chiral, and it is zero for achiral structures. An individual multipole in $\Psi^K_Q(U)$ must obey its site symmetry, e.g., symmetry expressed by the electric crystalline field [14]. Site symmetry is lower and more demanding than the crystal class (point group), with matching site symmetry and crystal class an exceptional case. In a crystalline sample, $\Psi^K_Q(U)$ can be different from zero if the crystal belongs to one of the 11 enantiomorphic crystal-classes, and it is zero for all other 21 crystal classes. A pair of enantiomers are physically and chemically indistinguishable in achiral settings. In the present case, the sample setting is made chiral by circularly polarized illumination. Taste distinguishes enantiomers of aspartame, with one sweet ("NutraSweet") and one bitter, while aroma distinguishes the enantiomers of carvone with one caraway aroma and one spearmint aroma. Biological properties can be dramatically different, as with thalidomide where one enantiomer in the racemic mixture is a sedative and the other a teratogenic agent. Chirality is found in all biological systems; left-handed amino acids, right-handed glucose, DNA strands and proteins.

Reported amplitudes (6) and (7) omit factors that determine relative magnitudes, in particular radial integrals of the position variable taken between the core state and valence state. By way of an example, consider the photon energy E tuned to an L-edge, 2p. The ratio of E1-E2 to E1-E1 amplitudes is then,

$$\Re(1, 2) = [(\alpha E)/(2\, a_o R_\infty)]\, (\langle 2p|r^2|4p\rangle / \langle 2p|r|3d\rangle),$$

where α, $a_o$ and $R_\infty$ are the fine structure constant, Bohr radius and the Rydberg unit of energy, respectively. In so far as hydrogenic forms of radial wavefunctions are appropriate for the photo-ejected 2p electron and empty 3d and 4p valence states $\langle 2p|r^2|4p\rangle / \langle 2p|r|3d\rangle = -(1.66\, a_o / Z_o)$ where $Z_o$ is the effective core charge seen by the jumping electron. M1 and E1 dipoles in $\Re(1, 1)$ have magnitudes $\mu_B$ and $(e\, a_o)$, respectively, where $\mu_B$ is the Bohr magneton. Using $\mu_B/(e\, a_o) = \alpha/2$ leads to,

$$\Re(1, 1) = (\alpha\, a_o\, \langle \lambda|\lambda'\rangle)/\langle 2p|r|3d\rangle.$$

In this expression, $\langle \lambda|\lambda'\rangle$ is the overlap of orbitals in the M1 event that possess the same angular momentum, because the magnetic moment operator is diagonal in this basis [15].

## V. MAGNETIC CIRCULAR DICHROISM

In light of the foregoing description of an E2 event in terms of h(*r*), defined in (2), it follows that an E2-E2 event of immediate can be described by a product of such functions. With a definition,

$$H^K_Q = \{h \otimes h'\}^K_Q, \qquad (8)$$

the rank possesses five values, $K = 0, 1, ..., 4$. The E2-E2 amplitude is then [11],

$$F(2, 2) = \sum_{K,Q} (-1)^{K+Q} H^K_{-Q} \Psi^K_Q(T), \qquad (9)$$

with,

$$\Psi^K_Q(T) = \sum_\mathbf{d} \langle T^K_Q \rangle_\mathbf{d}. \qquad (10)$$

Atomic multipoles $\langle \mathbf{T}^K \rangle$ are the subject of an Appendix, together with sum-rules for integrated intensities. The multipoles are parity-even and time-odd (magnetic) for *K* odd. In many magnetic materials site symmetries are different from those experienced in the paramagnetic phase, with the operator for time-reversal among the elements of symmetry [14]. For a crystalline sample, the electronic structure factor $\Psi^K_Q(T)$ is prescribed by one of the 122 magnetic crystal-classes.

On inserting **ε** and **κ** appropriate for a superchiral beam in the definition of h(*r*), and inserting **ε**′ and **κ** to generate h′(*s*) we find MCD has an amplitude,

$$F(2, 2) = (1/2) \sqrt{(1/10)}\, P_2 [\Psi^1_0(T) - 2\, \Psi^3_0(T)]. \qquad (11)$$

The magnitude of MCD is discussed in Appendix A. There one finds that the dipole $\langle \mathbf{T}^1 \rangle$ includes $\langle \mathbf{S} \rangle$ and $\langle \mathbf{L} \rangle$, while the octupole $\langle T^3_0 \rangle$ includes $\langle L_0 [5L_0^2 - 3L(L+1) + 1] \rangle$.

With regard to the relative magnitude of MCD we follow the previous discussion of NCD and calculate a ratio of E2-E2 and E1-E1 amplitudes. Sticking with the example of absorption at the L-edge,

$$\Re(2, 2) = [q(\langle 2p|r^2|4p \rangle / \langle 2p|r|3d \rangle)]^2,$$

with $q = E/(\hbar c)$.

## VI. DISCUSSION

We have completed an analysis of dichroic signals retrieved from samples illuminated by a superchiral standing wave that enhances signals with respect to measurements obtained using a circularly polarized beam [1, 2]. Like the Borrmann effect [6, 7, 8], the enhancement in question arises from the suppression of electric dipole-allowed transitions at the nodes. This aspect is vividly represented in our examination in Appendix B of enhancement factors available for dichroic signals using E1-E2, E1-M1 and E2-E2 absorption events. The benefit is a straightforward gain in signal strength because natural circular (NCD) and magnetic circular (MCD) dichroic signals are found to be precisely the same as the signals generated by a circularly polarized beam [11, 12, 13]. Previous studies of dichroic signals derived from a superchiral beam examined NCD alone [1-5]. NCD is here attributed to both E1-E2 and E1-M1 absorption events, and we added a study of MCD based on an E2-E2 event. In all cases, dichroic signals are expressed in terms of atomic multipoles that encapsulate electronic degrees of freedom in the ground-state. Atomic multipoles for MCD are gathered in Appendix A. Use of an effective wavevector is central to our method of working [9].

In summary, we have:

- Established an intimate connection between the Borrmann effect, twisted beams and superchiral beams [9].
- Shown that all three can be described by established spectroscopic theory by utilizing an effective wavevector.
- Highlighted the connection between optical and x-ray spectroscopies via the common language of atomic multipoles.
- Extended discussion of the use of superchiral beams to magnetic materials and electric quadrupole transitions.
- Interpreted the enhancement of NCD and MCD offered by a superchiral standing wave with respect to measurements obtained using a circularly polarized beam.

**Acknowledement** One of us (SWL) is grateful to Dr Michael Cochran for guidance with handling animated cartoons.

## APPENDIX A: E2-E2 MULTIPOLES

When the primary photon energy, E, is close to an atomic resonance with an energy $\Delta$ the scattering length approaches,

$$f \approx - (r_e \, \Xi \, F) \{\Delta/[E - \Delta + i\Gamma/2]\}, \tag{A1}$$

where F is the amplitude discussed in the main text. The resonance has a lifetime $\propto \hbar/\Gamma$, and the classical radius of an electron $r_e \approx 0.282 \, 10^{-12}$ cm is chosen as the unit of length. In a dimensionless pre-factor for an E2-E2 event,

$$\Xi \approx 2.64 \, \Delta^3 \, [\langle r^2 \rangle / a_o^2]^2 \, \aleph(l_c, l), \tag{A2}$$

$\Delta$ is measured in units of keV, $a_o$ is the Bohr radius, and $\langle r^2 \rangle = \langle l_c | r^2 | l \rangle$ is the radial integral for the core state and valence states that possess orbital angular momenta $l_c$ and $l$, respectively. The sum $(l_c + l)$ is even for the E2 absorption event [11, 16]. In due course, we consider two

special cases, namely, absorption at a K-edge ($l_c = 0$, $l = 2$) and an absorption event with $l_c = l$, e.g., 2p → 4p. For the remaining factor in (A2) one finds,

$$\aleph(s, d) = 1 \text{ and } \aleph(l, l) = (l\|L\|l)^2/[(2l - 1)(2l + 3)], \tag{A3}$$

with $(l\|L\|l) = \sqrt{\{l(l+1)(2l+1)\}}$.

Hydrogenic forms of radial wavefunctions for the photo-ejected 1s electron and empty 3d valence state yield $\langle 1s|r^2|3d \rangle = 1.73\,(a_o/Z_o)^2$ where $Z_o$ is the effective core charge. Likewise, $\langle 2p|r^2|4p \rangle = -7.90\,(a_o/Z_o)^2$.

An atomic multipole $\langle T^K \rangle$ is a sum of two operator types with $\langle \Omega^K \rangle$ composed exclusively of orbital angular momentum and $\langle \Phi^K \rangle$ more complicated by virtue of its dependence on both spin and orbital angular momentum. The actual relationship depends on the total angular momentum for the core state $J_c = l_c \pm 1/2$; one finds $\langle T^K \rangle = [\langle \Omega^K \rangle + \langle \Phi^K \rangle]$ for $J_c = (l_c + 1/2)$ and $\langle T^K \rangle = [\{l_c/(l_c + 1)\}\langle \Omega^K \rangle - \langle \Phi^K \rangle]$ for $J_c = (l_c - 1/2)$. This simple dependence on $J_c$ is correct for E1-E1 and E2-E2 events [11, 16, 17, 18, 19]. Multipoles and selections rules are different for the two events, of course. At the K-edge there is an explicit cancellation of $\langle \Omega^K \rangle$ from $\langle T^K \rangle$ with $J_c = (l_c - 1/2)$. Likewise, our expression for $\langle \Phi^K \rangle$ is zero under identical conditions, leaving $\langle T^K \rangle = \langle \Omega^K \rangle$ evaluated with $J_c = 1/2$.

In the general case,
$$\langle \Omega^K \rangle = \langle O^K \rangle \left[(l_c + 1)/\{(l\|L\|l)(2l_c + 1)\}\right] \sqrt{(2K+1)} \begin{Bmatrix} l_c & 2 & l \\ K & l & 2 \end{Bmatrix}, \tag{A4}$$

with a dipole $\langle O^1{}_0 \rangle = \langle L_0 \rangle$, and an octupole,

$$\langle O^3{}_0 \rangle = \langle L_0 [5L_0{}^2 - 3L(L+1) + 1] \rangle / [(l-1)(2l-1)(l+2)(2l+3)]^{1/2}, \tag{A5}$$

where the operator for orbital angular momentum $L_0 \equiv L_z$. The 6j-symbol in (A4) has the value $-(1/5)$ for $K = 1$ & 3 at the K-edge ($l_c = 0$, $l = 2$). For the other case of immediate interest,

$$\begin{Bmatrix} l & 2 & l \\ 1 & l & 2 \end{Bmatrix} = -[1/(l\|L\|l)]\sqrt{(3/10)},$$

$$\tag{A6}$$

$$\begin{Bmatrix} l & 2 & l \\ 3 & l & 2 \end{Bmatrix} = \quad [6/(l\|L\|l)][\{2(l-1)(l+2)\}/\{35(2l-1)(2l+3)\}]^{1/2}.$$

For $K = 3$ the angular momentum $l$ in the valence state must exceed 1.

The mixed spin and orbital multipole is derived from,

$$\langle \Phi^K \rangle = 2\,[\{l_c\,(l_c+1)\}/(2l_c+1)]^{1/2} \sum_b (2b+1)\,(-1)^b\, \langle \mathbf{B}^K(b) \rangle \begin{Bmatrix} 2 & l_c & l \\ 2 & l_c & l \\ K & 1 & b \end{Bmatrix}. \quad (A7)$$

Using the tensor product defined in (2) we have $\langle \mathbf{B}^K(b) \rangle = [\langle \{\mathbf{S} \otimes \Upsilon^b\}^K \rangle / (l\|\Upsilon^b\|l)]$ where $\mathbf{S}$ is the spin operator and $\Upsilon$ is a suitable spherical tensor operator built from $\mathbf{L}$. The 9j-symbol can be different from zero for integer $K + b$ odd, so $b$ is even for magnetic circular dichroism. The value $l_c = 0$ in the 9j-symbol is not allowed, meaning $\langle \Phi^K \rangle = 0$ at the K-edge. For the dipole allowed $b = 0$ & 2. One finds $\langle \mathbf{B}^1(0) \rangle = [\langle \mathbf{S} \rangle / \sqrt{(2l+1)}]$ and,

$$\langle B^1{}_0(2) \rangle = \sqrt{(2/3)}\,\langle \{L\,(L+1)\,S_0 - (3/2)\,(L_0\,\mathbf{S}\cdot\mathbf{L} + \mathbf{S}\cdot\mathbf{L}\,L_0)\} \rangle / (l\|\Upsilon^2\|l), \quad (A8)$$

with $(l\|\Upsilon^2\|l) = (l\|L\|l)\,[(2l-1)(2l+3)]^{1/2}$. 9j-symbols associated with $l_c = l$ are,

$$\begin{Bmatrix} 2 & l & l \\ 2 & l & l \\ 1 & 1 & 0 \end{Bmatrix} = \{(l\|L\|l)\,[10\,(2l+1)]^{1/2}\}^{-1},$$

$$\begin{Bmatrix} 2 & l & l \\ 2 & l & l \\ 1 & 1 & 2 \end{Bmatrix} = \{5\,(l\|L\|l)\,(l\|\Upsilon^2\|l)\}^{-1}\,[4\,l\,(l+1) - 9]. \quad (A9)$$

Octupole partners $\langle B^3{}_0(b) \rangle$ to (A5) are complicated quantities that we choose not to specify [16, 19].

### APPENDIX B: ENHANCEMENT FACTORS

Looking at the electron-photon interaction V, equ. (1), the E1 matrix element is linear in the electric field whereas the E2 matrix element contains a product of the electric field and the wavevector. Likewise, the M1 matrix element contains a product of the electric field and the wavevector [10, 11, 13]. With these facts in mind we proceed with a calculation of an enhancement factor for a superchiral beam that is an analogue of the dissymmetry factor used by Tang and Cohen [1, 2].

Electric fields of the secondary (back-scattered) and primary waves differ by a factor $\{\exp(i\phi)\sqrt{R}\}$, where R is the reflectivity of the mirror and $\phi$ is the all-important phase angle. Using a primary wave of unit amplitude, the standing-wave field possesses and effective electric field $\hat{E}_+ = [1 + \exp(i\phi)\sqrt{R}]$ that is the sole factor scaling the corresponding E1 matrix element.

Turning to E2 and M1 matrix elements we consider the product of the electric field and the wavevector, $\mathbf{q}$. For a unit amplitude primary wave, the product scales as $\mathbf{q}$ while the product scales as $\{\mathbf{q}'\exp(i\phi)\sqrt{R}\}$ for the secondary wave. In consequence, E2 and M1 matrix elements scale as $\hat{E}_- = [1 - \exp(i\phi)\sqrt{R}]$ on taking $\mathbf{q} = -\mathbf{q}'$ in the standing wave-field with near-perfect back-scattering.

Scattering amplitudes $\propto V^+V$ scale as $|\hat{E}_+|^2$ and Re. $[\hat{E}_+(\hat{E}_-)^*] = (1 - R)$ for E1-E1, and E1-E2 & E1-M1, respectively. We regard the ratios,

$$g(1, 2) = g(1, 1) = (1 - R)/|\hat{E}_+|^2, \qquad (B1)$$

as enhancement factors available from a superchiral photon beam for E1-E2 and E1-M1 dichroic signals. Equality of $g(1, 2)$ and $g(1, 1)$ was noted by Tang and Cohen [2]. Evidently, suppression of the E1-E1 event in (B1) is the sole factor in determining enhancement of parity-odd signals, and $g(1, 2) = g(1, 1) = [(1 + \sqrt{R})/(1 - \sqrt{R})]$ with $\phi = 180º$ [1, 2]. Saturation of enhancements factors is discussed by Choi and Cho [20].

Tang and Cohen [1, 2] do not examine E2-E2 dichroic signals for which we find a different value for g, namely,

$$g(2, 2) = \{|\hat{E}_-|^2/|\hat{E}_+|^2\}. \qquad (B2)$$

Setting $\phi = 180º$ in (B2) yields a superior enhancement $g(2, 2) = g(1, 1)^2$.

FIG. 1. Top panel; Cartesian coordinates (x, y, z) with the x-axis normal to the plane that contains the wavevectors **q** (primary) and **q**′ (secondary) with $\mathbf{q} \cdot \mathbf{q}' = 2q^2 \cos(2\theta)$. Bottom panel; wavevectors for the Borrmann case.

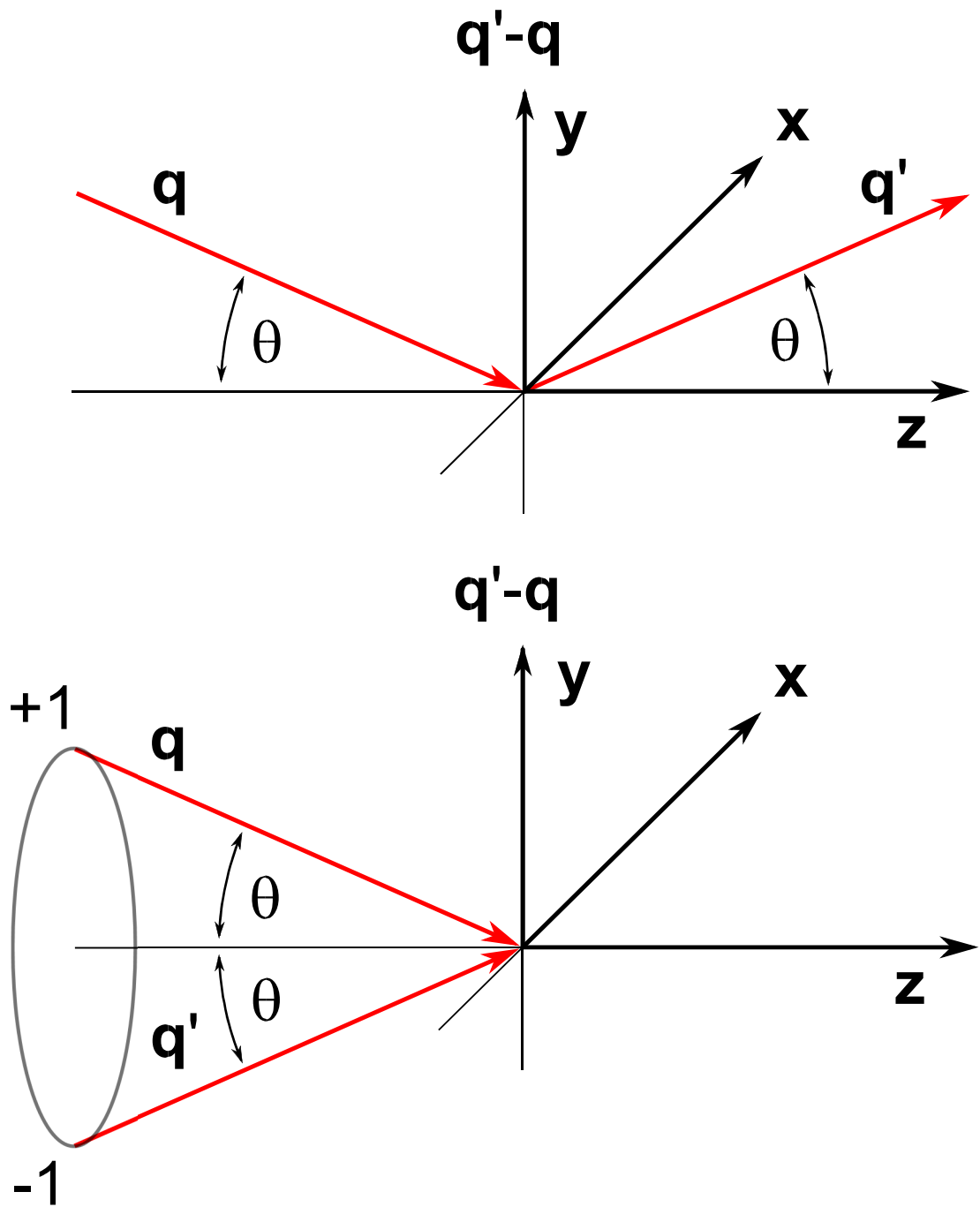

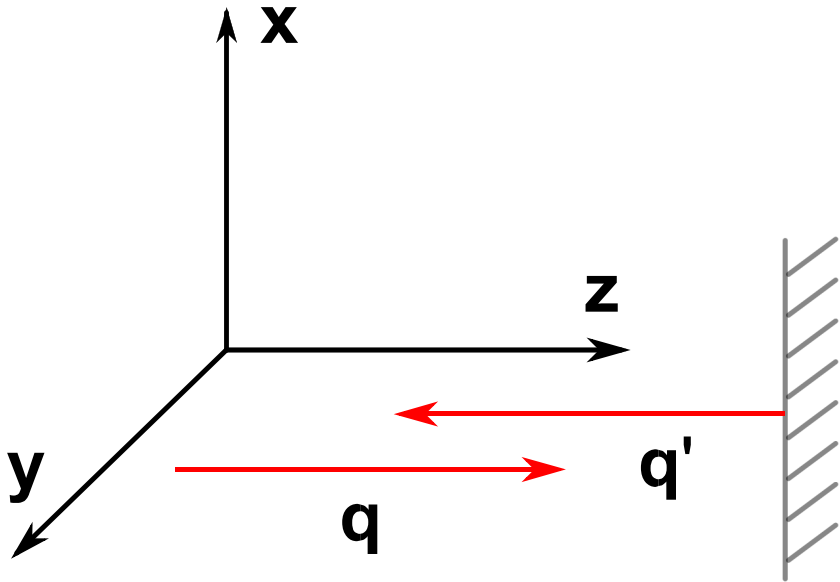

FIG. 2. Cartesian coordinates (x, y, z) adopted for a superchiral beam depicting the primary, **q**, and back-scattered, **q'** (mirror reflection), components. The effective wavevector **κ** is parallel to the z-axis and polarization vectors **ε** and **ε'** lie in the x-y plane.

FIG.3. Schematic representation of a superchiral standing wave-field, with additional representations in Supplementary Material. Counter-propagating circularly polarized beams of amplitudes of 0.3 and 0.7 form a standing wave with non-zero amplitude at the field nodes. The elliptical profile of the standing wave rotates about the z- axis with time, depicted by four example time-steps. Around the nodes (clarified by squares), the electric polarization vector twists sharply in comparison to the anti-nodes, indicative of high optical chirality. As the amplitudes of the forward- and counter-propagating waves approach equality, the optical chirality increases up to a fundamental limit and amplitude decreases [20]. Chiral optical effects exhibited by particles located at the nodes (depicted by spheres) are enhanced by this high optical chirality.

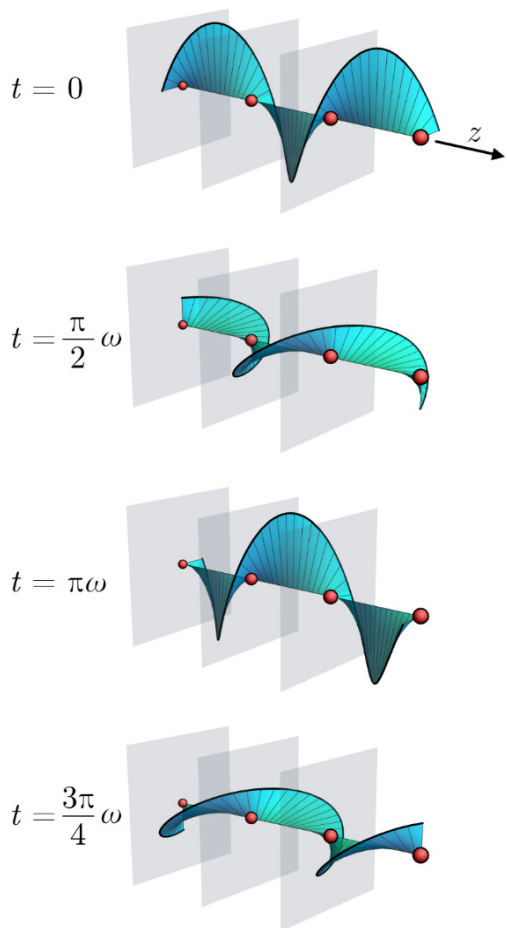